\documentclass[prb,showpacs,print,twocolumn]{revtex4}
\usepackage{graphicx}
\usepackage{dcolumn}
\usepackage{bm}
\usepackage{txfonts}
\begin{document}

\title{Dependence of electronic and optical properties on
a high-frequency field for carbon nanotubes}

\author{Wenhu Liao$^1$}
\author{Guanghui Zhou$^{1,3}$}
\email{ghzhou@hunnu.edu.cn}
\author {Kai-He Ding$^2$}

\affiliation{$^1$Department of Physics and Key Laboratory of
Educational Ministry for Low-Dimensional Quantum Structures and
Manipulation, Hunan Normal University, Changsha 410081,
China\footnote{Mailing address}}

\affiliation{$^2$Department of Physics and Electronic Science,
Changsha University of Science and Technology, Changsha 410076,
China}

\affiliation{$^3$International Center for Materials Physics, Chinese
Academy of Sciences, Shenyang 110015, China}

\date{\today}

\begin{abstract}
We study theoretically the electronic structure, transport and
optical properties for a zigzag single-wall carbon nanotube
connected to two normal conductor leads under the irradiation of an
external electromagnetic field at low temperatures, with particular
emphasis on the features of high-frequency response. Using the
standard nonequilibrium Green's function techniques, we examine the
time-averaged density of states, the conductivity, the dielectric
function and the electron energy loss spectra for the system with
photon polarization parallel with the tunneling current direction,
respectively. Through some numerical examples, it is shown that the
density of states is strongly dependent on the incident electron
energy, the strength and frequency of the applied field. For higher
electron energies in comparison with lead-nanotube coupling energy,
the system conductance decreases with increasing the field strength
and increases with increasing the field frequency respectively, and
shows some oscillation structures. Moreover, the optical functions
for the system have also a rich structure with the variation of
field frequency. It may demonstrate that this transport dependence
on the external field parameters can be used to give the energy
spectra information of carbon nanotubes and to detect the
high-frequency microwave irradiation.
\end{abstract}
\vspace{0.2cm}

\maketitle

\section{Introduction}
Nanoelectronics, or molecular electronics, have been proposed as the
alternative to silicon in future technical applications$^1$ and have
attracted much interest recently. Especially, carbon based
nanostructures, such as fullerene, graphene and nanotubes, are the
most interesting structures because of their excellent physical and
chemical properties. Carbon nanotubes (CNTs) can exhibit metallic or
semiconducting behavior with different diameter and chirality, and
therefore they will be promising candidates for the new
carbon-nanotube-based electronic devices, such as nanosensors,$^2$
electric batteries,$^3$ field-effect transistors,$^{4,5}$ Coulomb
blockade devices,$^{6,7}$ and field-emission displays.$^8$

The quantum transport properties of CNT systems have been
investigated experimentally by many authors.$^{9-11}$ The coherent
quantum wire behaviors of an individual single-walled carbon
nanotube (SWCNT)$^9$ between two contacts and the conductance
quantization of multi-walled CNTs$^{10}$ have been observed,
respectively. Moreover, Onac {\it et al.}$^{11}$ have used an
on-chip to detect the high-frequency noise signals generated by
quantum dot formed in a SWCNT with photon-assisted tunneling. This
result so far indicates no intrinsic frequency limitation due to the
CNTs themselves. Using the long-wavelength approximation, Mele {\it
et al.}$^{12}$ have studied the coherent one-photon and two-photon
electronic excitations for graphene sheets and SWCNTs. The optical
dielectric function of a finite SWCNT has also been studied early by
gradient approximation$^{13}$ and by first-principle
calculations,$^{14}$ respectively. Furthermore, the conductances for
microwave field irradiated multiwall CNT$^{15}$ and SWCNT$^{16}$
connected to two leads have been investigated by photon-assisted
transport model, in which the conductance oscillations evolve toward
a well-defined step structure and its sensitive dependance on the
field intensity have been expected.$^{16}$ This study has been
extended$^{17}$ to optical selection rule for SWCNT very recently.
The question thus arises whether SWCNTs could be useful as terahertz
(THz) detection, which is more important in the use of quantum
control to nanostructures. To the best of our acknowledge, this
important case for a driven SWCNT-based system has not been reported
previously.

In this paper, we present a theoretical investigation of the
electronic structure, transport and optical properties for a
metallic zigzag chiral SWCNT under a THz electromagnetic field (EMF)
irradiation since metallic CNTs are known to be one of the ideal
system exhibiting quantum mechanical nature of the electrical
transport.$^{15}$ The dependence of time-averaged electron density
of states (DOS), conductance, dielectric function ($\varepsilon$)
and electron energy loss spectrum (EELS) on the irradiation field
strength and frequency are demonstrated for the system,
respectively. Some different characteristics is obtained and the
results are compared with those for the similar physical systems in
the previous works.$^{9,13-17}$

The rest of the paper is organized as follows. The analytical
expressions of the time-averaged DOS, the conductance and the
optical functions are calculated starting from the system
Hamiltonian by NGF approach in Sec. II. Some numerical examples and
discussions for the results are demonstrated in Sec. III. Finally,
Sec. IV concludes the paper.

\section{Model and Formulism}
The NGF approach has been employed in last decades to study
transport problems involving variety of interactions beyond the
linear response regime.$^{18}$ Meir {\it et al.}$^{19}$ have derived
a formula for the current through a region of interacting electrons
using the nonequilibrium Keldysh formulism. Here we assume that the
external high-frequency field only causes a rigid shift in the
electron energy spectrum under the widely adopted adiabatic
approximation. Therefore, the Hamiltonian for an irradiated SWCNT
embedded between two normal conductor leads reads
\begin{eqnarray}
H&=&\sum_{q,\alpha\in{L/R}}\epsilon_{q,\alpha}d_{q,\alpha}^+d_{q,\alpha}
+\sum_{k}\epsilon_{k}c_{k}^+c_{k}\nonumber\\
&&+\sum_{q,\alpha\in{L/R},k}(V_{q,\alpha,k}d_{q,\alpha}^+c_{k}
+V^*_{q,\alpha,k}c_{k}^+d_{q,\alpha})\nonumber\\
&&+\sum_{k,k'}V_{0}\cos(\omega t)c_{k}^+c_{k'},
\end{eqnarray}
where the operator ${d^+_{q,\alpha}}$ ($d_{q,\alpha}$) creates
(annihilates) an electron with momentum $q$ in mode $\alpha$ in
either left (L) or right (R) lead, and $c_{k}^+$ ($c_{k}$) creates
(annihilates) an electron at the state $k$ of the CNT with energy
spectrum$^{13,20}$ $\epsilon_{k}$=$\pm\gamma[3+
2\cos(2k_{y}b)+4\cos(k_{y}b)\cos(3k_{x}a)]^{1/2}$, which is similar
as that of graphene though the boundary conditions of SWCNT are
different from graphene, especially in the circle direction. Where
$\gamma$ is the hopping integral of CNT in the tight-banding
approximattion, $a\approx$0.71{\AA} is the half length of in-plane
lattice constant with $b$=$\sqrt3a$, and $k_x$ and $k_y$ are
in-plane wavevectors of Dirac electron. The coupling between the
electrode leads and the CNT with strength $V_{q,\alpha,k}$ is
represented by the third term in Hamiltonian (1), and the last term
describes the electron-photon interaction in the CNT, where $V_{0}$
is the interaction strength under dipole approximation and $\omega$
the incident field frequency. Moreover, for simplicity, in the above
Hamiltonian we have neglected the spin and electron-electron
interactions although much attention has been paid to the
pseudospin-related effects in graphenes$^{21}$ and SWCNTs.$^{22}$

Now we employ the usually defined retarded and correlated Green's
Function (GF)$^{16,18,19}$ as $G_k^r(t_2,t_1)$=$\mp i\theta(\pm t_2
\mp t_1)\langle\{c_k(t_2),c_k^+(t_1)\}\rangle$ and
$G_k^<(t_2,t_1)$=$-i\langle\{c_k(t_1),c_k^+(t_2)\}\rangle$ to the
Hamiltonian. When the unperturbed GF $g^{r}(\epsilon)$ of the
nanotube is obtained, one can obtain GFs of $G^{r}(\epsilon)$ and
$G^<(\epsilon)$ from the Dyson equation
$G^r(\epsilon)$=$[(g^r(\epsilon))^{-1}-\Sigma^r(\epsilon)]^{-1}$ and
the Keldysh equation
$G^<(\epsilon)$=$G^r(\epsilon)\Sigma^<(\epsilon)G^a(\epsilon)$,
respectively. Furthermore, the self-energies are
$\Sigma^r(\epsilon)$=$(-i/2)[\Sigma^{L}(\epsilon)+\Sigma^R(\epsilon)]$
and $\Sigma^<(\epsilon)$=$(-i/2)[f_L(\epsilon)\Sigma^L(\epsilon)+
f_R(\epsilon)\Sigma^R(\epsilon)]$, where the linewidth function
$\Sigma^{\alpha}(\epsilon)$=$2\pi\sum_{q,k}V^*_{q,\alpha,k}
V_{q,\alpha,k}\delta(\epsilon-\epsilon_{q,\alpha})$ describes the
influence of the leads and $f_{\alpha}(\epsilon)$ is the Fermi
distribution function. In the wide-bandwidth
approximation,$^{15,16,18,19}$ $\Sigma^{\alpha}(\epsilon)$ is taken
to be independent of the energy and energy levels.

Furthermore, the GF of a honeycomb lattice (graphene) can be written as
\begin{equation}
g_{k}^{r}(\epsilon)=\frac{\epsilon/\gamma}{\epsilon^2-3-
2\cos(2k_{y}b)-4\cos(k_{y}b)\cos(3k_{x}a)}.
\end{equation}
For a $(n,0)$ chiral zigzag SWCNT, the periodic boundary
conditions$^{13,16}$ requires that the transverse wave vector $k_y$
supplies $k_yb$=$\pi j/n$ with good quantum numbers for the subbands
$j$=1$\sim$2n, and the longitudinal wave vector $k_x$ is confined in
the first Brillouin zone $-\pi/(3a)$$<$$k_x$$<$$\pi/(3a)$. Applying
Fourier transform to $g_k^r(\epsilon)$ one obtains the unperturbed
GF in the real space
\begin{widetext}
\begin{eqnarray}
g^r(\epsilon,l,m)=\frac{6ab\epsilon}{4\pi^2\gamma}\sum_{j=1}^{2n}
\int_{-\pi/(3a)}^{\pi/(3a)}dk_x\frac{e^{-3ik_xla}e^{-i\pi
jm/n}}{\epsilon^2-3- 2\cos(2\pi j/n)-4\cos(\pi j/n)\cos(3k_xa)},
\end{eqnarray}
\end{widetext}
where $l$ and $m$ are respectively the length and width index in
real space, once the integration over $k_x$ is done in Eq. (3), one
can obtain the analytical expression of $G^{r}(\epsilon)$ from Dyson
equation. Therefore, the real part of the zero-bias limit linear
conductance$^{15-17}$ can be derived as
\begin{eqnarray}
\sigma_1(V_0,\omega)&=&\frac{2e^2}{h}\int d\epsilon
\sum_{\mu}J_{\mu}^2\bigg(\frac{V_0}{\hbar\omega}\bigg)\nonumber\\
&&\times\bigg(-\frac{\partial f(\epsilon )}{\partial \epsilon
}\bigg) \Sigma^{L}G^{r}(\epsilon)\Sigma^{R}G^{a}(\epsilon),
\end{eqnarray}
where $J_{\mu}(x)$ is the $\mu$th-order Bessel function of the first
kind. The imaginary part of conductance, $\sigma_2(V_0,\omega)$, can
be calculated from $\sigma_1(V_0,\omega)$ by Kramers-Kronig
transformation$^{14,17}$
\begin{eqnarray}
\sigma_1(V_0,\omega)=1+\frac{4}{\pi}\textbf{P}\int^\infty_0
d\omega^{'}\frac{\omega^{'}\sigma_2(V_0,\omega^{'})}{{\omega^{'}}^2-\omega^2},
\end{eqnarray}
where \textbf{P} denotes the principal value of the integral. From
the complex dielectric function $\varepsilon(V_0,\omega)$=1+4$\pi
i\sigma(V_0,\omega)/\omega$ with $\sigma(V_0,\omega)$=
$\sigma_1(V_0,\omega)$+$\sigma_2(V_0,\omega)$, in principle one can
obtain all other linear optical properties such as reflectivity,
absorption spectrum and EELS by $-\Im
m[\varepsilon(V_0,\omega)]^{-1}$ at the long-wavelength
limit,$^{13,14}$ where $\Im m$ represents the imaginary part of the
quantity. The effective electron DOS will be$^{16}$
\begin{equation}
D(\epsilon)=\sum_{\mu}\bigg|J_{\mu}\bigg(\frac{V_0}{\hbar\omega}\bigg)\bigg|^2
D_0(\epsilon-\mu\hbar\omega),
\end{equation}
In the tight-banding approximation, the DOS of a $(n,0)$ CNT in the
absence of perturbing potential reads$^{22}$
\begin{eqnarray}
D_0(\epsilon)&=&\frac{2}{3\pi^2}\Im m \sum_{j=1}^{2n}i(
\epsilon+i0^+)\times \big\{8[1+\cos(2\pi j/n)]\nonumber\\
&-&[(\epsilon+i0^+)^2-2\cos(2\pi j/n)-3]^2\big\}^{-1/2}.
\end{eqnarray}

It should be pointed out that Eq. (6) can be interpreted as
following. Photon absorption ($\mu$$>$0) or emission ($\mu$$<$0) can
be viewed as creating an effective electron DOS at sideband energies
$\epsilon_{\mu}$=$\mu\hbar\omega$ with a probability
$|J_{\mu}(V_0/\hbar\omega)|^2$. One notes that the multi-photon
processes will be suppressed more or less in the cases of
$V_0/(\hbar\omega)$$\gg$1, and the enough strong EM field
($V_0/(\hbar\omega)$$\gg$1 cases) will split sharp singularities in
the electron spectrum except for its affection on the electron
distribution of the CNT.$^{23}$ Therefore, in this work we
concentrate on the cases of the relatively high-frequency response
under moderate irradiation field strength for the system.

\section{Results and Discussions}
In the following, we present some numerical examples of the
calculated normalized DOS, linear conductance and optical properties
for an external field irradiated (12,0) SWCNT. The tight-banding
approximation hopping integral $\gamma$=2.75 eV is taken to be the
unit of energy, the temperature$^{24}$ and the effective `speed of
light' are $k_B$T=0.01 and $v_F$=8.88$\times10^5$ m/s, respectively.
In our numerical calculation, the geometry index $l$ and $m$ in Eq.
(3) are selected as 12 and 24 for a SWCNT system, that is, the
diameter and length of the CNT is about 1 nm and 2.6 nm,
respectively. In the wide-band approximation,$^{15,16,18,19}$ we
assume that $\Sigma^{L}$=$\Sigma^{R}$=$\Sigma$=0.001$\gamma$ as the
CNT-lead coupling parameter for simplicity. The above selected
parameters for the system are reachable in the present
experiments.$^{9-11}$
\begin{figure}
\center
\includegraphics[width=3.0in]{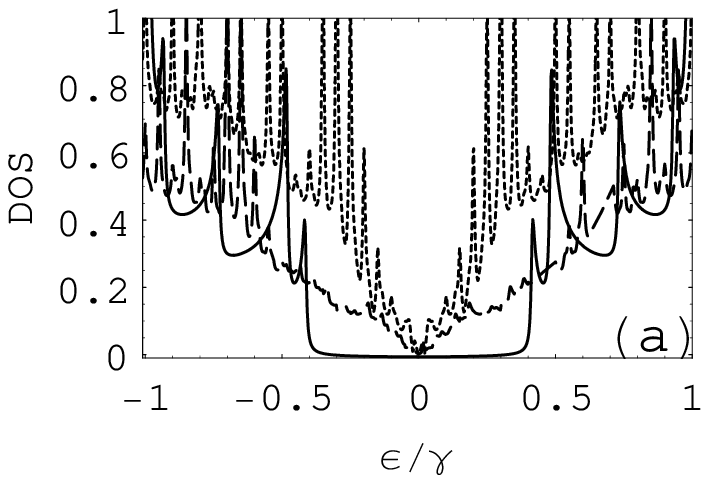}
\includegraphics[width=3.0in]{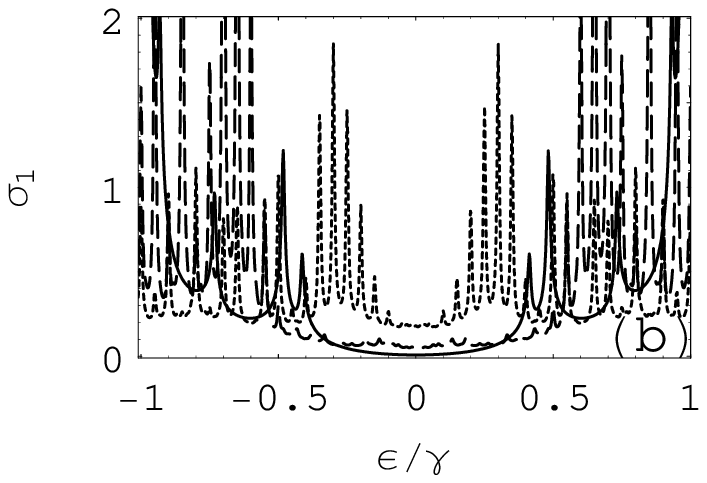}
\caption{The dependence of (a) time-averaged DOS and (b) conductance
on the incident electron energy $\epsilon$ with a field frequency
$\nu$=33.25 THz for three different field strengths, where solid
line [down-shifted 0.1 for comparison in (a)] for $V_0$=0, dashed
line for $V_0$=1.1 eV and dotted line [up-shifted 0.1 for comparison
in (b)] for $V_0$=2.2 eV, respectively.}
\end{figure}

Figure 1 shows the calculated time-averaged DOS (in arbitrary units)
and linear conductance $\sigma_1$ (in units of $\sigma_0$=2$e^2/h$)
as a function of the incident electron energy $\epsilon$ (in units
of $\gamma$) with a fixed field frequency
$\nu$=$\omega/(2\pi)$=33.25 THz for different field strength $V_0$.
In the absence of irradiation field, as shown by the solid line in
Fig. 1(a), the DOS presents a series of original peaks at energies
$\pm$0.40, $\pm$0.50, $\pm0.75$ and $\pm0.95$, respectively. A wide
van Hove pseudogap is observed wherever the DOS is not zero.
However, when an irradiation field of strength $V_0$=1.1 eV with
photon energy $h\nu$=0.05$\gamma$ is applied, as shown by the dashed
line in Fig. 1(a), except for a series of resonance peaks in low
energy range of -0.40$\gamma$$<$$\epsilon$$<$0.40$\gamma$, there are
some resonance peaks at $\pm$0.55, $\pm$0.60, $\pm$0.65, $\pm$0.70,
$\pm$0.80, $\pm$0.85 and $\pm$0.95$\gamma$, respectively. These
additional peaks are mainly attributed to the processes of 1-photon
and 3-photon processes. Moreover, in the case of field strength
$V_0$=2.20 eV [see the doted line in Fig. 1(a)] with the same field
frequency, due to the 2-photon, 4-photon and 6-photon processes, the
system DOS presents a much richer structure except for a remarkably
changed van Hove pseudogap. This result is quantitatively different
from that in Ref. 16 (see, e.g., Fig. 2 in Ref. 16) where no
quantitative explanation about the electronic structures was given.
The sensitive dependence of DOS on the irradiation field strength
can be physically interpreted as following. With a certain photon
frequency, the variation of the field strength should be followed by
different photon sideband processes as can be verified from Eq. (6).
Furthermore, the electronic structures of an irradiated $(10,0)$
zigzag SWCNT (semiconducting type) have also been plotted with the
same parameters (not shown here), the results are similar except for
a wider van Hove pseudogap is observed.

The corresponding real part conductance $\sigma_1$ to Fig. 1(a) is
illustrated in Fig. 1(b). In the absence of external filed, as shown
by the solid line in Fig. 1(b), $\sigma_1$ presents a peak at the
four original resonance states and $\sigma_1$ is nearly 0 in the
pseudogap as expected. Wherever, under the irradiation of a field
with strength of $V_0$=1.10 eV, as shown by the dashed line in Fig.
1(b), the system conductance presents a series of additional peaks
at $\pm$0.55, $\pm$0.60, $\pm$0.65, $\pm$0.70, $\pm$0.75 and
$\pm$0.85$\gamma$, respectively. When the field strength increases
to 2.20 eV, one can find several additional conductance peaks at
$\pm$0.15, $\pm$0.20, $\pm$0.25 and $\pm$0.30$\gamma$ in the low
energy range [as shown by the doted line in Fig. 1(b)] owing to
multi-photon processes, while the conductance peaks in the higher
energy range are unchanged.
\begin{figure}
\center
\includegraphics[width=3.0in]{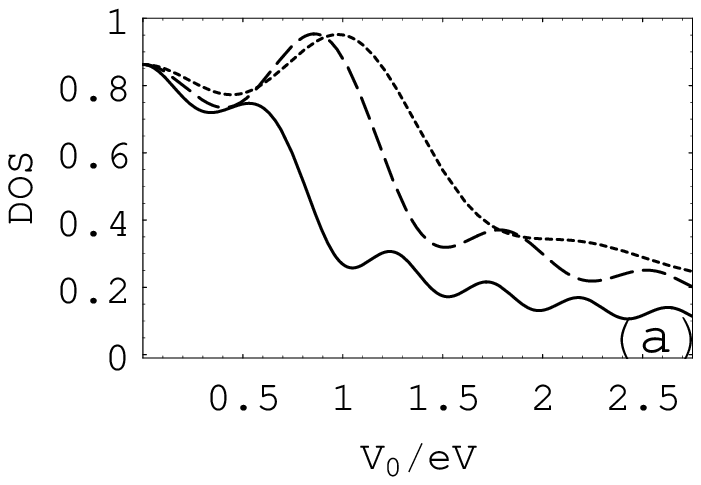}
\includegraphics[width=3.0in]{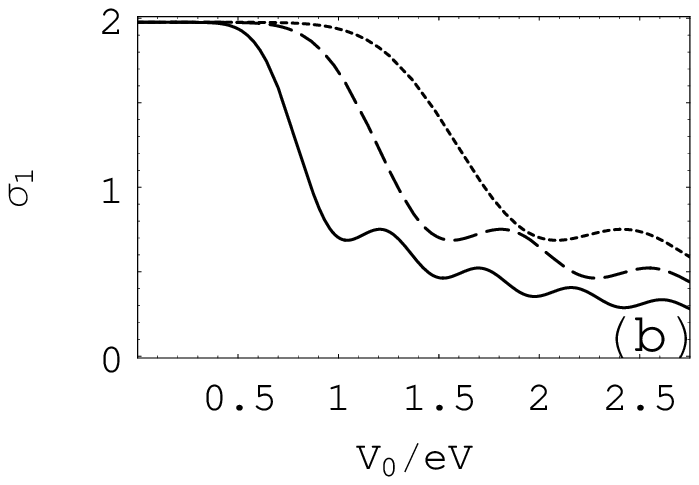}
\caption{The dependence of (a) time-averaged DOS and (b) conductance
on the field strength with a fixed energy $\epsilon$=2.0 eV for
three different field frequencies, where solid line for $\nu$=33.25
THz, dashed line for $\nu$=49.9 THz and dotted line for $\nu$=66.5
THz, respectively.}
\end{figure}

In what follows, we show the influences of the external field
parameters on the electronic structure and transport of the system
for a fixed incident electron energy $\epsilon$=8$\gamma$/11 (2.0
eV). Fig. 2 depicts the time-averaged DOS and $\sigma_1$ as a
function of the irradiation field strength ($\sim V_0$). In the case
of a relative low field frequency $\nu$=33.25 THz, as shown by the
solid line in Fig. 2(a), the DOS decreases slowly with a oscillating
periodicity about 0.5 eV other than a step at 0.50$<$$V_0$$<$1.0 eV.
When the field frequency increases to $\nu$=49.9 THz (dashed line)
and $\nu$=66.5 THz (dotted line) as shown in Fig. 2(a), the
oscillating periodicity seems to become larger except for a peak at
0.9 eV and 1.0 eV, respectively.

Figure 2(b) illustrates the behavior of $\sigma_1$ as a function of
the field strength for three different field frequencies. In the
case of low frequency $\nu$=33.25 THz, as shown in the solid line,
$\sigma_1$ presents a downgrade step-like structure with oscillating
periodicity about 0.5 eV similar to that of the DOS [see Fig. 2(a)].
When the field frequency increases to $\nu$=49.9 THz or $\nu$=66.5
THz, as shown in the dashed line or doted line of Fig. 2(b), the
oscillating periodicity becomes larger. As mentioned above, due to
the case of higher frequency for the system we have considered, the
results here are quantitatively different from those in Ref. 16
(see, e.g., Figs. 4 and 6 in Ref. 16) where well-defined up-going
step structures for the conductance have been predicted only for
very low ($\nu$$<$3.3 THz) field frequencies with zero incident
electrons energy and temperature. It should be pointed out that the
results here are qualitatively different those in Ref. 15 (see,
e.g., Figs. 2 and 3 in Ref. 15) where a nearly universal down-going
normalized resistance with the microwave power and
frequency-irrelevant normalized resistance at liquid helium
temperature have been observed. Moreover, the electronic structures
and conductance of a longer zigzag SWCNT have been plotted (not
shown here) with the same all other parameters, the results are
qualitatively similar as expected.
\begin{figure}
\center
\includegraphics[width=3.0in]{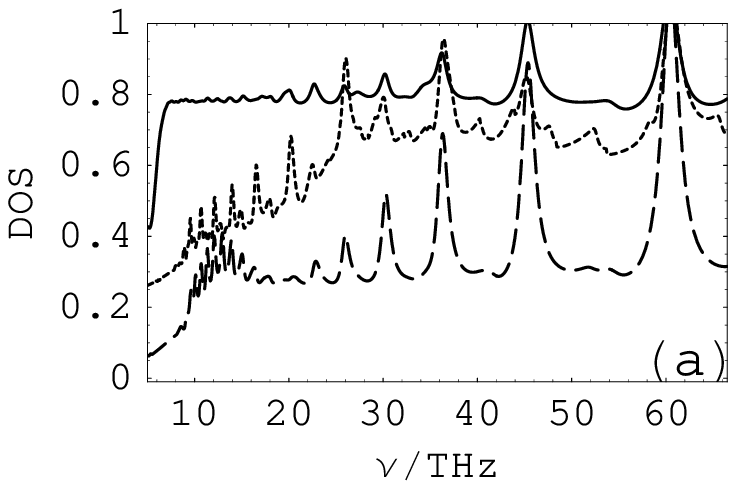}
\includegraphics[width=3.0in]{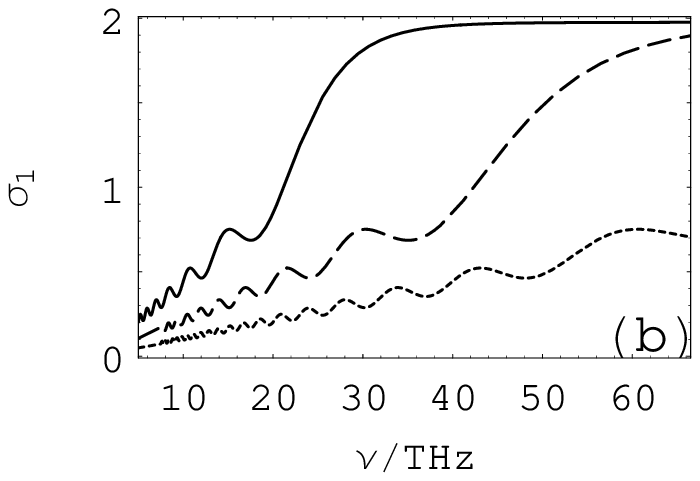}
\caption{The dependence of (a) time-averaged DOS and (b) conductance
on the field frequency with a fixed incident electrons energy
$\epsilon$=2.0 eV for three different field strengths, where solid
line for $V_0$=0.55 eV, dashed line for $V_0$=1.1 eV and dotted line
[up-shifted 0.2 for comparison in (a)] for $V_0$=2.2 eV,
respectively.}
\end{figure}

Figure 3 demonstrates the calculated time-averaged DOS and
conductance $\sigma_1$ as a function of the irradiation field
frequency $\nu$ (in units of THz) with a fixed incident electron
energy $\epsilon$=2.0 eV (same as in Figure 2) for three different
field strengths. As shown in Fig. 3(a), one notices that in all case
of field strength there is a series of peaks at frequencies
$\nu$=10, 12, 14, 16.5, 20, 22.5, 26, 30, 36, 45.5 and 60 THz,
respectively. Owing to the difference of the field strengths, these
peaks should result from different photons processes as mentioned
above and the characteristics will be manifested in the conductance
more or less. The behavior of the time-averaged normalized
conductance $\sigma_1$ versus the field frequency for different
strengths is illustrated in Fig. 3(b). In the case of smaller
strength $V_0$=0.55 eV (see the solid line), $\sigma_1$ presents a
rapidly oscillatory up-going structure with the increase of field
frequency, and reaching almost 2.0$\sigma_0$ as field frequency
reaches 36.5 THz. However, when $V_0$ increases to 1.10 eV, as shown
the dashed line in Fig. 3(b), $\sigma_1$ up-goes slowly, only two
obvious upgrade step-like structures at 20 and 30 THz, respectively.
Furthermore, even a much more slowly oscillatory structure is shown
by the doted line in Fig. 3(b) for the case of lager $V_0$=2.20 eV.
It should point out that the series of resonance peaks are due to
multi-photon processes. The above results are quantitatively
different from those in Ref. 16 (see, e.g., Fig. 5 in Ref. 16) where
a nearly independence of field frequency in the range of
0$<$$\nu$$<$0.005$\gamma$ and a higher frequency
0.005$\gamma$$<$$\nu$$<$0.05$\gamma$ oscillatory response of the
conductance have been predicted.

\begin{figure}
\center
\includegraphics[width=3.0in]{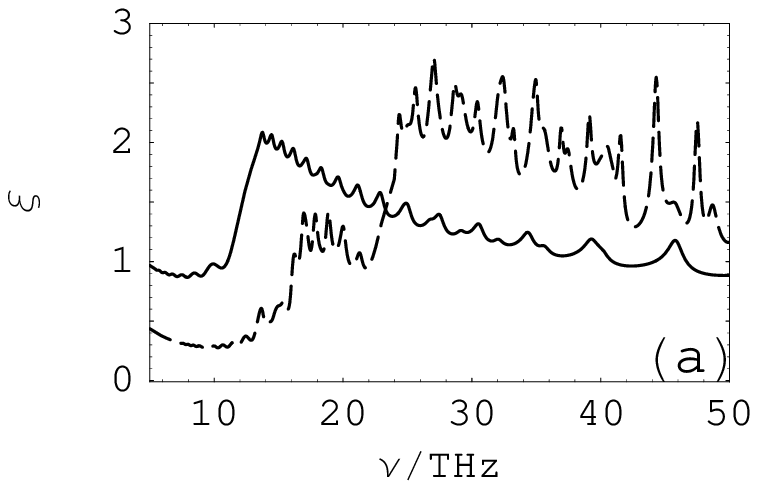}
\includegraphics[width=3.0in]{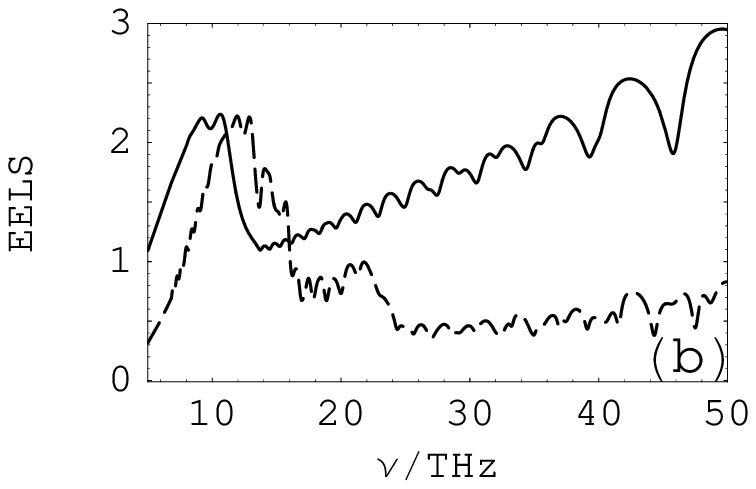}
\caption{The dependence of (a) dielectric function
$\varepsilon(\nu)$ and (b) electron energy loss spectra (EELS) on
the field frequency with a fixed energy $\epsilon$=2.75 meV for two
different field strengths, where solid line [up-shifted 0.5 for
comparison in (a) and (b)] for $V_0$=1.1 eV and dashed line for
$V_0$=2.2 eV, respectively.}
\end{figure}
Next, we present numerically the dielectric function
$\varepsilon(\nu)$ and EELS versus the irradiation field frequency
with a fixed incident electron energy (selected as 2.75 meV for
comparison with the results in Refs. 14-17) for two different field
strengths, and further discuss the basic optical
properties$^{13-17}$ of the system. As shown in Fig. 4(a),
$\varepsilon(\nu)$ in dashed line, one notices a broad optical
absorption frequency strip at the height of 1.0 and 2.0 in the
ranges of 16.5$<$$\nu$$<$23.5 THz and 24$<$$\nu$$<$30 THz
respectively, with down-going oscillation structures in the range of
30$<$$\nu$$<$50 THz for the case of $V_0$=2.2 eV. Wherever, a peak
of near 2.0 height in the vicinity of 13.5 THz with down-going
oscillation structures in the higher frequency range for $V_0$=1.1
eV (see the solid line). The characteristics of the
$\varepsilon(\nu)$ are different from the results in Refs. 13, 14
and 17 since different parameters we have chosen, although the
principal features of Fig. 8 in Ref. 14 have been revealed.
Corresponding to the $\varepsilon(\nu)$ as show in Fig. 4(a), the
EELS is illustrated in Fig. 4(b). For the case of $V_0$=2.2 eV EELS
presents a richer structure as expected since it is related to
$\varepsilon(\nu)$ by $-\Im m[\varepsilon(\nu)]^{-1}$. Except for
the two broad peaks in the neighbor of 10 and 12 THz for $V_0$=1.1
and 2.2 eV, one notes a optical absorption strip at the height of
1.5 and 0.8 in the ranges of 13.5$<$$\nu$$<$16 THz and
17$<$$\nu$$<$23 THz respectively, with oscillation structures around
0.5 in the range of 24$<$$\nu$$<$50 THz for the case of $V_0$=2.2 eV
[see the dashed line in Fig. 4(b)]. Since different nanotubes always
have a dominating plasmon mode$^{13}$ at $\sim$6 eV, therefore, all
the above features should rely on multi-photon processes. In short,
all the electron DOS, conductance, dielectric function and EELS of
CNTs system are sensitive to the strength and the frequency of the
external field, which may be applied to the detection of the
external THz field and the energy spectra information of carbon
nanotubes.

\section{Conclusion}
In summary, using the NGF method, we have investigated theoretically
the electronic structure, transport and optical properties of a
metallic-typed zigzag single-wall carbon nanotube connected with two
normal leads under the irradiation of an external high-frequency
(THz) electromagnetic field at low temperatures. Using the standard
nonequilibrium Green's function techniques, we examine the
time-averaged density of states, conductance, dielectric function
and electron energy loss spectrum for the system with photon
polarization parallel with the tunneling current direction,
respectively. It is demonstrated that, by analyzing some numerical
examples, the density of states shows a strong dependance on the
incident electron energy and external field parameters. For high
irradiation energies in comparison with CNT-lead coupling, the
system conductance decreases with the field strength while increases
with the field frequency followed by some oscillation structures.
For the lower incident electron energy, the dielectric function and
electron energy loss spectrum are consistent with the results in
Refs. 13, 14 and 17 due to multi-photon processes. In all cases the
system transport and optical properties are sensitive to the
parameters of the external field, which may be utilized to the
detection of high-frequency irradiation and the energy spectra
information of carbon nanotubes. However, the experimental
observation for these effects and further theoretical investigations
on the system with impurity, spin or electron-electron interactions
and the armchair chiral carbon nanotubes are worthy to be carried
out.

\begin{acknowledgements}
This work was supported by National Natural Science Foundation of
China (Grant No. 10574042), Specialized Research Fund for the
Doctoral Program of Higher Education of China (Grant No.
20060542002) and Hunan Provincial Natural Science Foundation of
China (Grant No. 06JJ2097).
\end{acknowledgements}

\end{document}